\documentclass[journal]{IEEEtran}
\usepackage{graphicx}
\usepackage{tipa}
\usepackage{bbm}
\usepackage{mathrsfs}
\usepackage{amsfonts}
\usepackage{cite,url,subfigure,epsfig,graphicx}
\usepackage{amssymb,amsmath}
\usepackage{indentfirst}
\usepackage{algorithmic}
\usepackage{algorithm}
\usepackage{pdfpages}
\usepackage{cite,url,subfigure,epsfig,graphicx}
\usepackage{times,verbatim,amsfonts,amsmath,color}

\IEEEoverridecommandlockouts
\makeatletter
\def\ps@headings{\def\@oddhead{\mbox{}\scriptsize\rightmark \hfil \thepage}\def\@evenhead{\scriptsize\thepage \hfil \leftmark\mbox{}}\def\@oddfoot{}\def\@evenfoot{}}
\makeatother 

\begin{document}

\title{Futuristic Intelligent Transportation System}

\author{Yilong Hui,~\IEEEmembership{Member,~IEEE,}
Zhou Su,~\IEEEmembership{Senior Member,~IEEE,}
Tom H. Luan,~\IEEEmembership{Senior Member,~IEEE},
and Nan Cheng,~\IEEEmembership{Member,~IEEE}
%Xiao Xiao,~\IEEEmembership{Member,~IEEE}

}

\maketitle

\begin{abstract}
%\boldmath
The emerging autonomous vehicles (AVs) will inevitably revolutionize the transportation systems. This is because of a key feature of AVs; instead of being managed by human drivers as the conventional vehicles, AVs are of the complete capability to manage the driving by  themselves. As a result, the futuristic intelligent transportation system (FITS) can be a centrally managed and optimized system with the fully coordinated driving of vehicles, which is impossible by the current transportation systems controlled by humans. In this article, we envision the operation of such FITS when AVs, advanced vehicular networks (VANETs) and artificial intelligence (AI) are adopted. Specifically, we first develop the autonomous vehicular networks (AVNs) based on the advanced development of AVs and heterogeneous vehicular communication technologies to achieve global data collection and real-time data sharing. With this network architecture, we then integrate AVNs and AI based on the intelligent digital twin (IDT) to design the FITS with the target of setting up an accurate and efficient global traffic scheduling system. After that, compared with the conventional schemes, a customized path planning case is studied to evaluate the performance of the proposed FITS. Finally, we highlight the emerging issues related to the FITS for future research.

\end{abstract}



\IEEEpeerreviewmaketitle

%----------------------------------------------------------------------------------

\section{Introduction}\label{sec:1}

\begin{figure}[t!]
\centering
\includegraphics[width=8cm]{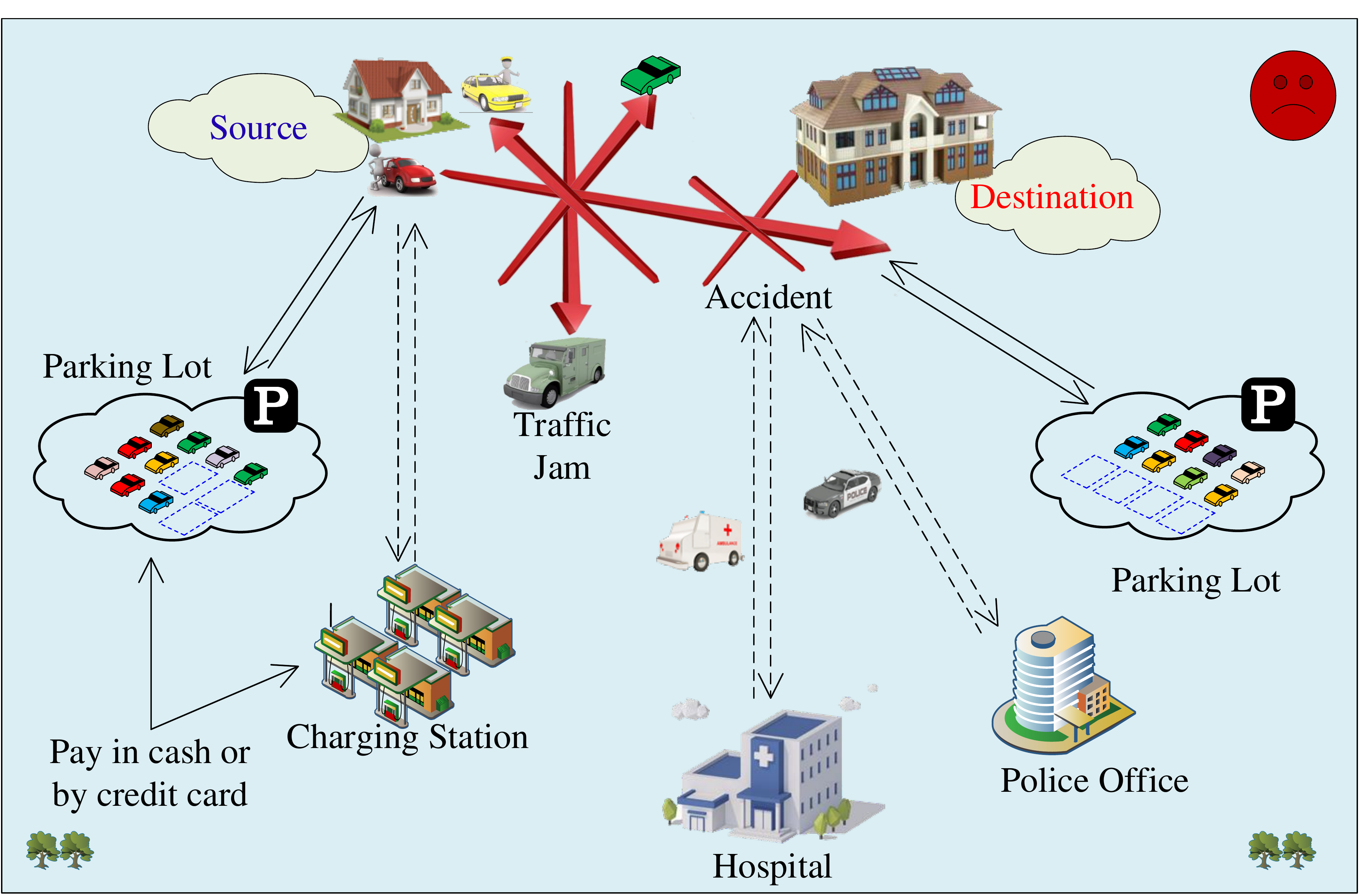}
\vspace{0.0cm}
\caption{The current ITS.}
\label{fig.1}
\end{figure}

With the rapid development of urbanization, the percentage of population residing in cities is envisioned to rise to 70\% in 2050 \cite{Rizvi2018aspire}. As a major means of transportation in cities, commute by vehicles is an essential part in our daily life. To this end, vehicular networks (VANETs) have attracted a lot of attention with the target of developing intelligent transportation system (ITS) and making our trip more convenient and enjoyable \cite{Silva2019information}. In ITS, vehicles and roadside operating units are wirelessly connected using vehicular communication technologies so as to provide drivers the real-time traffic information and assist them to make driving decisions. Despite the efforts on ITS, with vehicles controlled by human being, ITS is only an information system with following issues:

\textbf{Troublesome and dangerous work:} Controlled by driver, the driving actions of a vehicle is decided by the driver's personal experience and subjective consciousness. In addition, the drivers in a boring driving environment may be distracted by other things (e.g., phone ringing, scenery, etc.) along the way. As reported by the Association for Safe International Road Travel \cite{national}, the reason that causes accidents is mainly attributable to the inappropriate operations of drivers. Apart from safety problem, as shown in Fig.~\ref{fig.1}, drivers have to do a lot of labor intensive and time consuming works (e.g., park, charge and repair their vehicles) in the ITS and thus fail to obtain the comfortable travel experience.

\textbf{Inefficient scheduling system:} The ITS can provide information and recommend driving decisions to drivers. However, not all the drivers drive according to the recommended decisions. In addition, the road conditions provided by the ITS, in general, are the events that have happened. This phenomenon mainly stems from the fact that the traffic status is affected by the decisions of all the drivers, where the topology dynamically changes with the high-speed mobility of vehicles.
    Consequently, it is difficult to schedule and manage the traffic in a global way.
%    The rapid increase in the number of vehicles is making the problems of traffic worse. These problems, not only waste people's time, but also result in serious energy consumption and environmental pollution.

 \textbf{Low quality of experience:} With various travel intentions, different drivers have different driving requirements. For example, some drivers may pay attention to the driving time while the other drivers may care about the driving costs. However, the driving decisions that can be recommended by the ITS to each driver are limited, where the diverse driving requirements of drivers can not be satisfied. Furthermore, the recommended decisions in the ITS are not absolutely correct due to the lack of precise management and global control. As a result, the drivers may suffer from a low quality of experience (QoE) because the ITS fails to meet their diverse requirements.

Fortunately, the rapid development of autonomous driving technology makes it possible for revolutionizing the ITS. It is reported that all Tesla cars shipped in 2019 are equipped with full autonomous driving capability.
%In addition, Baidu has lunched Apollo Go Robotaxi, the
%self-driving taxis services, to the public in Beijing in
%October 2020.
Embedded with various sensors, computing devices and communication units, the controllable autonomous vehicles (AVs) can understand the surrounding traffic environment and make accurate driving decisions by themselves \cite{Bhat2018tools}. In this way, drivers are released from getting into boring and dangerous works, where the real-time data perception and precise vehicle control can significantly improve the driving safety. Furthermore, by integrating VANETs with artificial intelligence (AI), the AVs can be collaboratively managed with the target of achieving the global scheduling system and satisfying the diverse driving requirements. With these promising advantages, autonomous driving will inevitably revolutionize the operation mode of the traditional transportation system, namely, the ITS.

%With The Integration Of Vehicular Networks And Autonomous Driving, On The Other Hand, The City Traffic Has Shown New Features. First, As Avs Need To Sense And Understand The Traffic Environment, Heavy Computation Tasks Raised By The Perception Process Need To Be Executed During Driving. Second, In Order To Control And Manage All The Avs In A Global Way, The Information Including The Collected Traffic Information, The Scheduling Commands And So Forth, Need To Be Transmitted In Real Time. To This End, It Is Necessary To Cache The Useful Data Properly To Reduce The Data Transmission Delay. Third, As Humans Are No Longer Involved In The Driving Process, On One Hand, Passengers Pay More Attention To The Driving Requirements Than Before. The Electronic Finance System That Can Be Used For Managing The Costs/Benefits Of All The Avs Needs To Be Designed To Satisfy Users' Requirements. Being Faced With These New Features, How To Integrate The Vehicular Networks And Avs To Revolutionize The Current Icts And Establish A Comprehensive Perception, Real-Time Decision-Making And Global Scheduling System To Provide Passengers With Safe, Efficient And Comfortable Travel Experience Therefore Becomes A Challenge.

%\begin{figure*}[t]
%\centering
%\includegraphics[width=17.1cm]{fig/2.pdf}
%\vspace{0.0cm}
%\caption{An example of the application of vehicular networks in city transportation.}
%\label{fig.2}
%\end{figure*}

In this article, we represent the initial step toward the futuristic ITS (FITS) by jointly considering the available vehicular communication technologies, AI and the features of AVs. We first develop the autonomous vehicular networks (AVNs) based on the rapidly development of AVs and the advanced vehicular communication technologies to support globe data perception and real-time data sharing. Then, we integrate the AVNs and AI to design the FITS which is an accurate and efficient global traffic scheduling and management system. After that, a case of customized path planning is studied to evaluate the performance of the proposed framework. Finally, we discuss the research issues to highlight the future research directions.
\begin{figure}[t!]
\centering
\includegraphics[width=8.1cm]{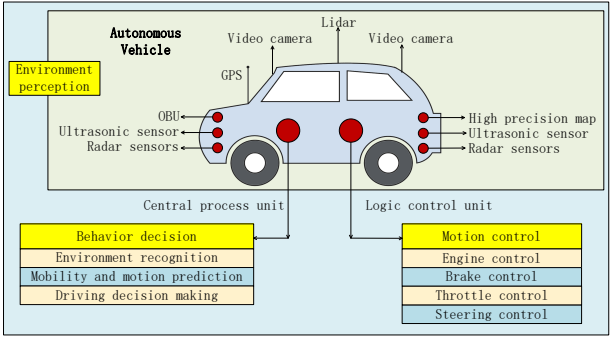}
\vspace{0.0cm}
\caption{The principles of autonomous driving.}
\label{fig.3}
\end{figure}

%
%The rest of this paper is organized as follows. In Section~\ref{sec:2}, we give a brief review of the AVs and the advanced vehicular communication technologies. The network architecture of AVNs is designed in Section~\ref{sec:3}. In Section~\ref{sec:4}, we introduce the proposed F-ICTS with the integration of AVNs and AI. In Section~\ref{sec:5}, we study a case to evaluate the proposed F-ICTS, followed by the conclusions in Section~\ref{sec:7}.
%----------------------------------------------------------------------------------

\section{Background of AVs and HetVNETs}\label{sec:2}

\begin{figure*}[t]
\centering
\includegraphics[width=14.1cm]{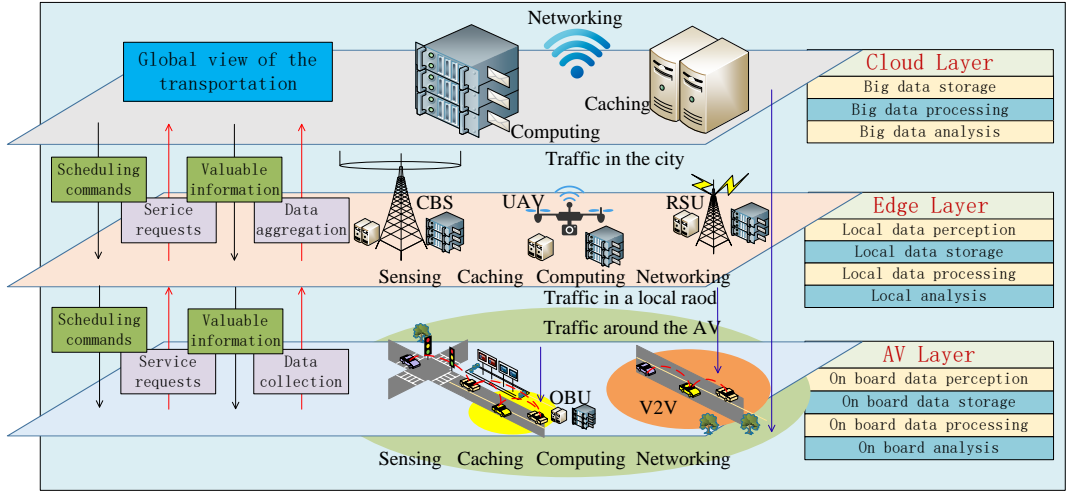}
\vspace{0.0cm}
\caption{The network architecture of the AVNs.}
\label{fig.5}
\end{figure*}

\subsection{AVs} \label{sec:21}

In order to perceive the driving environment and make driving decisions like humans, as shown in Fig.~\ref{fig.3}, an AV typically has the following components \cite{Liu2018intelligent}.
\begin{itemize}
\item \textbf{Sensing devices}: AVs, as a type of robots, are equipped with various sensors and devices such as lidar, radar sensors, video cameras, global position system (GPS) and inertial measurement unit (IMU). With these sensors and devices, each AV can understand the traffic environment and achieve a controllable driving.

%
%    Here, we briefly review the functions of the on-board sensors. 1) Lidar: Lidar is usually equipped on the roof of AVs. With high-speed rotation, the lidar can be used to identify the outline of objects around the AV in real time to generate high-resolution road maps. 2) Ultrasonic radars: According to the time difference between the delivered and received ultrasonics, the distance between the objects and the AV can be calculated. 3) Radar sensors: Radar sensors are charge of collecting the information around an AV, such as distance, velocity, and angle, by transmitting the millimeter wave and receiving the signal reflected by objects. 4) Video cameras: Camera is mainly used to collect images to estimate relative distance and speed of objects based on their moving mode. In addition, camera can also identify traffic signs and signals on the road to ensure that an AV's operation is strictly observed. 5) GPS and IMU: Global position system (GPS) is used to realize the localization of AVs. As the multipath problem of GPS signal will cause noise interference, the localization accuracy is low when purely relying on the GPS. To this end, inertial measurement unit (IMU) is typically used in conjunction with GPS to enhance the localization accuracy. With the GPS receiver and IMU, AVs become aware of their accurate mobile information, such as position, speed, directions, acceleration, turn rate, inclination and so forth.

\item \textbf{Storage device}: An on-board unit (OBU) is embedded in each AV to cache valuable information. On one hand, the private information of the AV can be cached in the OBU, such as the interests of the owner. On the other hand, the useful information, such as high precision maps, can be cached to facilitate the autonomous driving.

   % By comparing and mapping the environmental information perceived by the sensors with the detailed information labeled in high precision maps, the precise position of the AV in the city can be obtained. In addition, the detailed priori information provided by the high precision maps, such as the width of the road segment, the curvature of the road, the location of the road signs, etc., can be used to verify the perception results and assist the decision making.

\item \textbf{Processing device}: The central processing unit (CPU) is able to integrate the data sensed from surrounding environment and the useful data cached in OBU to make driving decisions.
%    To be specific, the CPU of an AV can recognize, predict and track objects in its surroundings by analyzing and computing the available data. Based on the analysis of the data, the CPU obtains the knowledge to determine the driving decisions and sends the corresponding decisions to the control unit.

 \item \textbf{Control device}: The logic control unit (LCU) completes real-time control of the onboard hardware devices to achieve safely autonomous driving.
%     In specific, the LCU is mainly used to execute the control operations, namely, output the control behavior to the brake, the throttle, the steering wheel etc., after receiving the driving decisions published by the CPU.
\end{itemize}

With these components, the autonomous driving process of an AV can be divided into the following
steps.

\begin{itemize}
\item \textbf{Environment perception}: The AV uses a group of sensors to sense the surrounding static objects (e.g., road signs and number of lanes) and dynamic objects (e.g., pedestrians, animals and other moving AVs), where the collected data are used to support driving decisions.

\item \textbf{Behavior decision}: Behavior decision is the process that the CPU makes driving decisions and generates the commands to control the AV's motion by jointly analyzing the perceived traffic environment information and the AV's current driving state.

\item \textbf{Motion control}: According to the driving decisions made by the CPU, the LCU then transmits the corresponding control commands to engines, brakes, wheels and so forth through wired signals with the target of realizing safe and precise motion control.
\end{itemize}

\subsection{HetVNETs}\label{sec:22}

With the development of space-air-ground integrated networks (SAGINs), a variety of communication technologies can form heterogeneous vehicular networks (HetVNETs) \cite{Cheng2020acompre}.
\begin{itemize}
\item \textbf{Satellite communication}: With a globe communication coverage, satellite communication is able to provide the network services for AVs which are out of the coverage of other vehicular communication infrastructures \cite{MacHardy2018v2x}.
%     As a consequence, satellite communications is usually integrated with other vehicular communication technologies to support seamless coverage and efficient resource
%management \cite{Cheng2020acompre}.

%A satellite network usually provides services for navigation, Earth observation, emergency rescue, and communication relaying. Based on the altitude, satellites can be
%categorized into geostationary orbit (GSO), medium Earth orbit (MEO), and low Earth orbit (LEO)
%satellites. Satellite networks have the advantages
%of:1) Large coverage (3 GSO satellites or a constellation of LEO satellites, e.g., Iridium composed of 77 LEO satellites can provide the
%global coverage).
%multi-layer satellite networks are combined.
%The advantages of diff erent orbits will complement
%each other's shortcomings to meet the quality
%of service (QoS) requirements of different users.
%$//$Space-Air-Ground IoT Network and
%Related Key Technologies

\item \textbf{DSRC}: Based on IEEE 802.11p, the dedicated short range communication (DSRC) is designed to provide network services to AVs \cite{Hui2020reservation}. To this end, a group of roadside units (RSUs) can be deployed along roads so that each AV can connect with its nearby RSU to obtain the requested information.
%    In addition, each AV has an OBU to cache information based on their interests and can share the cached information with its nearby AVs using vehicle to vehicle (V2V) communication.

\item \textbf{Cellular networks}: The cellular networks can be regraded as the long term evolution (LTE), 4G and 5G, which are advocated by the Third Generation Partnership Project (3GPP), to provide AVs with vehicle-to-everything (V2X) services \cite{Peng2018vehicular}. Using the cellular base stations (CBS), AVs can obtain network access services by connecting the cellular networks.

\item \textbf{UAVs}: Unmanned aerial vehicles (UAVs), which are equipped with dedicated sensors and communication devices, can undertake various vehicular services to facilitate the autonomous driving \cite{Shi2018drone}. With flexible mobility, UAVs can be treated as flying vehicular nodes to enhance the communication performance of the RSUs or OBUs.
\end{itemize}

%----------------------------------------

%----------------------------------------

\section{AVNs}\label{sec:3}

 %----------------------------------------
In this section, we design the AVNs by considering the features of AVs and the HetVNETs.

%----------------------------------------

\subsection{Network Architecture of AVNs}\label{sec:31}

As shown in Fig.~3, the architecture of the network consists of
three layers, which are AV layer, edge layer and cloud layer.
%The three layers complement each other to lay the foundation for efficient management of city traffic.

\textbf{AV layer:} This layer consists of all the AVs. Equipped with advanced sensors and OBU, each AV can sense the surrounding environment, cache valuable information and execute computing tasks for autonomous driving. Additionally, the perceived information can be shared among a group of AVs to make the driving decisions more accurately. However, caused by the perception range of sensors, the decisions made by AVs can only make driving decisions based on limited information.
%    On the other hand, AVs can connect with the communication devices deployed at the edge layer for requesting vehicular services using the V2I communications. For example, an AV can connect with the RSU for requesting the information about the optimal parking lot.

\textbf{Edge layer:}
  This layer is composed of edge infrastructures with caching, communication and computing abilities, such as RSUs, CBSs and UAVs. The RSUs and CBSs are fixed nodes while the UAVs in the contrast have high flexibility. The devices deployed at the edge layer are closer to the AVs than the cloud layer. Consequently, the edge layer can support local services with low latency. On the other hand, the edge devices can pre-process the data that need to be transmitted to the cloud layer, where the data traffic in the cloud can be reduced.

%  In this way, the devices deployed at the edge layer act as agents for assisting the cloud layer to gather the data generated by AVs and the data related to the city traffic in the local area.

%  Similarly, the devices in the edge layer can also serve as relays to distribute the valuable information published by the cloud layer to AVs. For example, edge devices can deliver the commands to the AVs in their communication coverage to charge the local traffic after receiving the schedule commands from the cloud layer.

 \textbf{Cloud layer:}
The cloud layer has more powerful caching and computing abilities than the AV layer and edge layer. On one hand, the historical data related to the traffic can be cached in this layer. For example, the information of an AV user can be cached in cloud server so that the travel habit of the user can be further estimated. On the other hand, the cloud layer has the global view of the traffic. It can obtain the information generated by the dynamic traffic and the AVs to make scheduling decisions and manage the traffic as a whole.

\subsection{Features of AVNs Enabled Transportation}\label{sec:33}

In this subsection, we analyze the features of AVNs enabled transportation to pave the way for developing the FITS.

\textbf{Global data perception:} With the advanced sensors and devices, the traffic environment around an AV can be efficiently collected. Furthermore, with the assist of AVNs, the data generated by the AVs together with the data generated by the traffic in the road sections can be collected by the infrastructures deployed at the edge layer to achieve the traffic data perception in a global way.

\textbf{Real time data transmission:}
With the support of the AVNs, the network coverage of the city can be realized and the traffic data can be shared among all the communication devices and infrastructures. On one hand, an AV can connect to the nearby AVs to deliver warning information or share driving information using V2V communication. On the other hand, the AVNs consist of various communication technologies can communicate with AVs to obtain real-time data sharing.

%s with different transmission performance. With different data transmission requests in diverse traffic environments, how to select configure the transmission parameters and select the optimal access network needs to be further studied.

\textbf{Multilayer data computing:}
In the AVNs, all the devices in the three layers have the computing ability and can be used to complete different tasks. For the AV layer, each AV can compute task by itself or execute task with other AVs collaboratively \cite{Su2018distributed}. Furthermore, the devices in the edge layer can execute computing tasks locally. If an AV has a task that cannot be executed by itself, the AV can offload the task to the nearby edge devices. Although the AVs and edge devices can assist the transportation with computing services, the resources that they have cannot satisfy the heavy computing tasks generated by the traffic. Therefore, some tasks need to be completed by the cloud server which can provide rich resources.

\textbf{Distributed data caching:}
The data with various types and values are cached in different devices in the AVNs. Specifically, the data about the AVs and their owners can be cached in the cloud server to develop insurance and diagnostic plans. The information related to geographic location can be distributed in the devices at the edge layer. For example, the high precision maps of all the roads in the city can be cached in the cloud server and a part of them can be cached in the edge devices and AVs selectively. Besides, the data related to the dynamic traffic are also distributed in different devices to facilitate the real time traffic scheduling. In this way, the retrieve time of the cached data can be reduced as some of them can be cached at the edge of the network.

%\textbf{Efficient data utilization:}
%The data can be efficiently utilized for facilitating the transportation with the support of AVNs. The data of AVs and their owners, which are cached in the cloud server, can be used to establish insurance and equipment replacement services. In addition, the data related to the travel habits of AV owners can be utilized to forecast and schedule the traffic flow. For the data collected by edge devices, they can be used to adjust the traffic flow and the driving speeds of AVs. As for the cloud server, it can obtain the valuable information to manage the traffic of all the road segments by analyzing the data gathered from edge devices. Moreover, the filtered data cached in the cloud server can be used to simulate the newly proposed algorithms or to train the deep learning models.

\section{FITS}\label{sec:4}
\begin{figure*}[t]
\centering
\includegraphics[width=13.1cm]{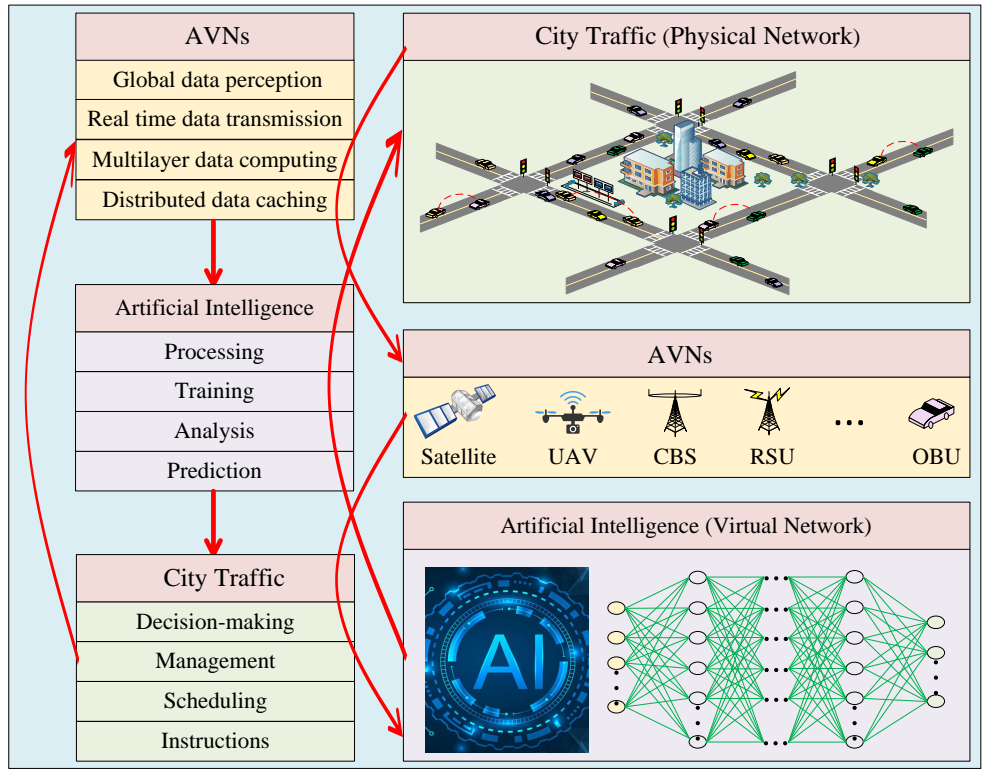}
\vspace{0.0cm}
\caption{The integration of AVNs and AI to support FITS.}
\label{fig.4}
\end{figure*}
According to the features of the AVNs enabled transportation,
%we can know that the following strategies are acquired to achieve a real-time,
%accurate and efficient global traffic scheduling and management
%system. 1) With diverse traffic environments, how to process and fuse the collected data to make accurately driving decisions; 2) With different data transmission requests, how to configure the transmission parameters and select the optimal access network needs to be further studied; 3) With various computation intensive tasks, how to divide the tasks with a number of subtasks and allocated to them to different devices deployed at different layers to improve the task computing efficiency; 4) How to optimally cache the data with various types
%and values to reduce the data transmission delay and improve the QoE of the data requester; 5) How to make full use of the data to promote the sustainable development of the transportation system. By jointly considering the requested strategies,
in this section, as the architecture shown in Fig.~4, we briefly introduce the process of combining AVNs and AI. With the AVNs, the data generated by the transportation (i.e., physical network) can be efficiently collected, shared, computed and cached. Then, based on the collected data, a virtual network can be developed by constructing the intelligent digital twin (IDT) of the physical network \cite{Zhao2020Intelligent}. In the virtual network, the data can be processed, analyzed, and analyzed to support various vehicular applications by using well-designed AI-based algorithms. With the AI-based algorithms, the traffic management and scheduling strategies and schemes can be obtained to facilitate the transportation system in the physical network. Based on the combination of AVNs and AI, we then design the FITS which includes the following systems, as shown in Fig.~\ref{fig.5}.

\textbf{Information collection system:}
The information collection system is able to manage all the data and information related to the transportation, including data generation, sensing, collection and utilization. These data can be collected to achieve different goals. With the analysis of the data by using AI-based algorithms, 1) the information of each AV and its owner cached in the cloud server can be used to make scheduling plans. For example, machine learning can be designed to detect the misbehavior of AVs \cite{Gyawali2020machine}; 2) the information of the components equipped on each AV, such as the lifetime of various on-board sensors, can be used to ensure the safety of the AV and develop insurance and
diagnostic plans; 3) the environmental information around each AV can be utilized to enable safe and efficient autonomous driving; 4) the data generated by the traffic of each road segment can be explored to forecast and schedule the traffic flow. Besides, compared with the high simulation cost in the actual system, the valuable data can be used to simulate and test new algorithms or train learning models. With the collected information, the transportation system then has a global view of the traffic. By using the global information to train AI algorithms, the cloud layer can make full use of the available resources in the FITS, such as charging station and parking lot.

 \textbf{AV control system:} The AV control system refers to the full automatic control process of an AV. In this process, each AV uses the AI-based algorithms to make driving decisions based on the collected data from the surrounding environment and the information received from the neighboring AVs or edge devices. With these decisions, there is no need for human intervention in driving and stopping operations as the motion of each AV can be controlled accurately and efficiently. For example, deep learning can significantly improve the AV's ability to
detect and classify the surrounding objects, thus facilitating
the fusion of data obtained by different sensors. With the AV control system, the accurate motion control of each AV can free drivers from getting into boring driving and significantly reduce the probability of accidents.

\textbf{Intelligent driving system:}
Based on the information obtained from AVs and the edge devices in the AVNs, the cloud server is in charge of the intelligent driving system. Specifically, by using the AI-based algorithms to analyze the collected data, the cloud layer can forecast the city traffic and dynamically decides a driving speed which will be assigned to the edge devices in the given road segment. Based on the assigned driving speed, the edge devices then control the AVs that drive on this road segment. For the AVs that join in this road segment, it keeps the driving speed as the same as others. On the contrary, for the AV that intends to change its lane, it needs to adjust its driving speed and follows the given speed in the new road segment. Obviously, with the traffic control of each road segment, the probability of traffic jams in each road segment can be reduced.

\textbf{Intelligent traffic scheduling system:}
With the intelligent driving system, the traffic in each road segment can be managed efficiently. We then introduce the intelligent traffic scheduling system to schedule the traffic in a global way. In the intelligent traffic scheduling system, the cloud layer can use AI-based algorithms to predict the traffic flow of each road section and then regulate the direction of traffic flow on each road to break the layout of roads so as to make flexible use of road resources. For example, with the efficient scheduling, traffic lights will no longer be needed to manage traffic at intersections \cite{Zhang2018vehicular}. Besides, rewards or compensations can be further used to dispatch traffic flow. Specifically, according to AI-based data analysis in the cloud layer, the system can give different rewards or compensations to different road sections. For example, the system can charge high prices on congested road sections and give rewards on non-congested road sections. In this way, some AVs will be attracted by rewards and drive to non-congested road sections, even if the non-congested road sections need to drive a longer distance.
According to the precise management of traffic in different road segments and intersections, the system can also develop real-time plans to deal with emergencies. For example, to make sure the rescue AVs (e.g., police cars and ambulances) to drive with a high-speed and barrier-free environment, the system will provide the best driving route to them and notify the AVs in the road segments associated with the selected route in advance. With the traffic scheduling, the transportation becomes smart, where the probability of traffic jams can be reduced and the traffic efficiency can be significantly improved.

\textbf{Electronic trading system:}
In order to minimize human participation, it is necessary to establish an electronic trading system to enable AVs to charge/pay fees according to their behaviors in the FITS. With the electronic trading system, each person, which intends to join in the FITS, has a virtual currency account (VCA) to complete the transactions with various types. Each AV owner needs to deposit enough money in advance in its VCA to make sure that its AV can interact with other devices successfully. For the people who do not have AVs, the VCA is also developed for each of them as they may travel by using the public transportation. After completing the transaction, the related fees will be automatically transferred among all the participants, thus reducing the number of brakes on AVs and avoiding the wrong payment caused by human error. As the resources in city transportation such as parking spaces and charge stations are limited and dynamically changing, AI-based algorithms can be used to dynamically price these resources. For the owner of an AV, when the AV charges, occupies the parking space or executes other consumption applications, the related fees will be deducted automatically from the owner's VCA based on the ID of the AV. For the people who use public transport, AI-based algorithms can be used to train their biological features so that they can pay for services more convenient. In this way, people in the FITS do not need cash or credit cards to pay for any transactions. Thus, it can greatly enhance people's travel experience and improve the utilization rate of AVs.
\begin{figure}[t]
\centering
\includegraphics[width=8.1cm]{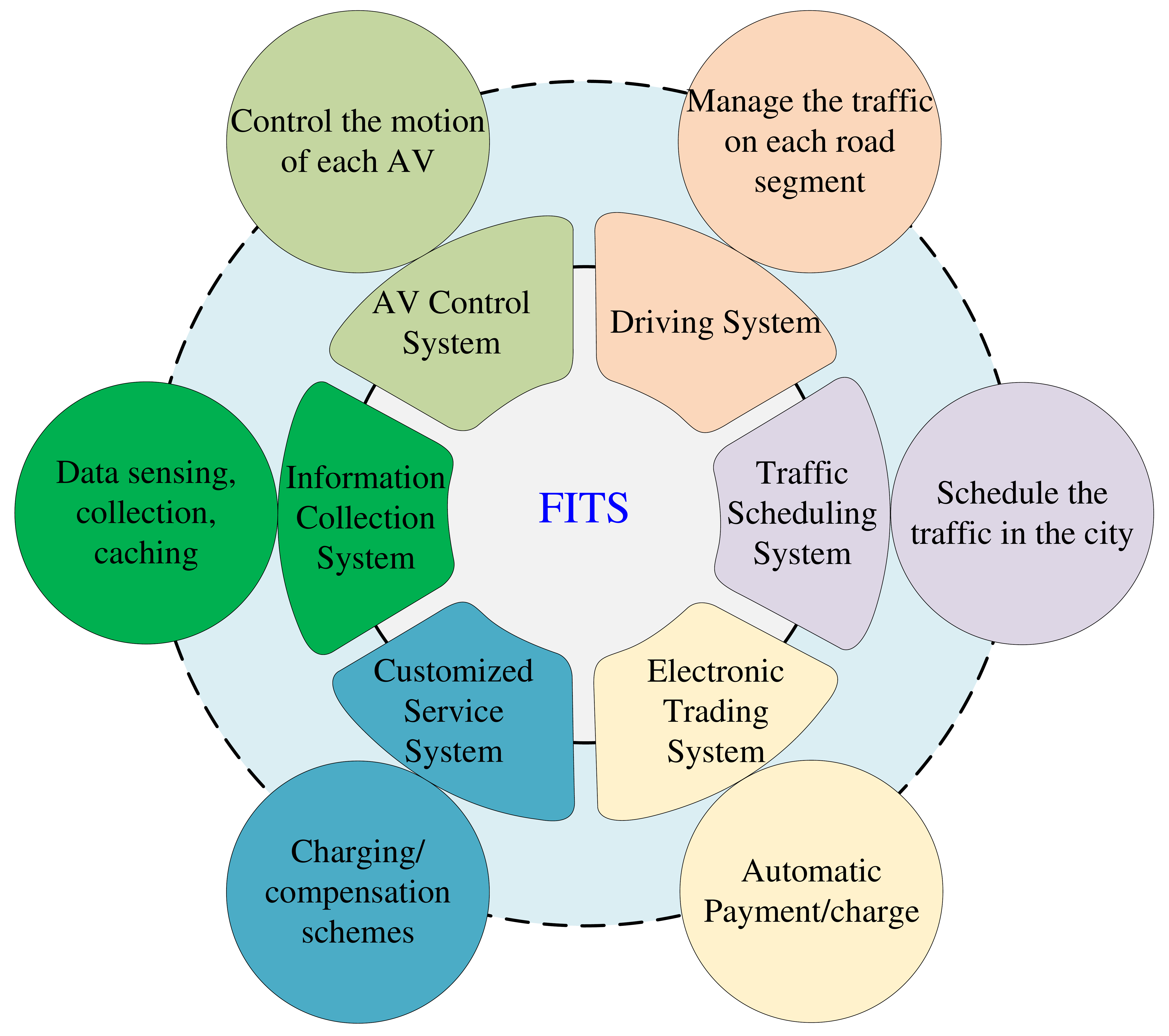}
\vspace{0.0cm}
\caption{The architecture of FITS.}
\label{fig.5}
\end{figure}

 \textbf{Customized service system:}
In the fully controlled traffic environment, passengers no longer need to focus on the driving process and thus pay more attention to the driving requirements. Generally, different people have different driving requirements (e.g., driving time and costs) even for the same trip. Therefore, we also need to establish the customized service system in the FITS to satisfy the needs of different users. For example, when an AV intends to park with low cost while does not care about the driving distance, the customized service system then provides the optimal choice to the AV for parking using the AI-based algorithms. In the aspect of traffic dispatching, the FITS can charge/compensate the cost to reduce/increase the travel time of users. By designing AI-based algorithms to analyze the collected data, the customized service system can set different charging/compensation schemes for users with different driving requirements. Specifically, if a user plans to reduce the travel time by selecting the shortest path and the fastest speed, it has to pay the system for the driving request. Conversely, the corresponding compensation will be paid the users who are willing to increase their travel time. In addition, according to the requirements of users, AI-based algorithms also can be explored to customize their driving routes.
Obviously, the customized service system can lubricate the city traffic scheduling and enhance the travel experience of each user.

\section{Case Study}\label{sec:5}

 In this section, a case of customized path planning is studied to demonstrate the
efficiency of the proposed FITS. We first introduce the scenario setting, followed
by the result analysis.

\subsection{Scenario Setting}

\label{sec:51} Consider a general scenario that an area has 100 road
sections, where the distance of each road section is randomly generated between 1km and 10km. In addition, we set the maximum speed (i.e., $v_{\max}$) and the maximum density (i.e., $k_{\max}$) of the AVs to be 110km/h and 80veh/km, respectively. In this way, we can obtain the velocity that an AV drives on each road section by $v=v_{\max}(1-\frac{k}{k_{\max}})$, where $k$ is the density of AVs in this road section \cite{Hui2019agame}. By dividing the distance by the speed, the cloud layer can collect the driving time to determine the compensation price for each road section, where the price changes from -1 to 1. Then the cloud layer delivers the driving time and compensation price to the edge layer which broadcasts these information to all the AVs. After receiving the driving time and compensation price of each road section, each AV uses the Q-learning algorithm to customize the optimal driving path to obtain maximum utility  based on its preference for compensation price and driving time, where the preference varies from 0 to 1 to weight the above two factors.

\subsection{Simulation Result}
With the aforementioned  scenario, we compare the proposal with the following conventional schemes: 1) Shortest driving time (SDT): The AV only considers the driving time to obtain the optimal driving path; 2) Maximum driving compensation (MDC): The AV only considers the compensation price to plan its driving path; 3) Minimum number of sections (MNS): The AV selects the path which has the least number of road sections. Fig. 6 is the utility of the AV by changing the preference of the AV for the compensation price and driving time. From this figure, we can see that the proposal can lead to the highest utility to the AV compared with the other conventional schemes. The reasons are as follows. For the SDT and MDC, they only consider one of the two factors (i.e., the driving time and the compensation price) to determine the optimal driving path, where the other factor is not considered to guarantee the utility of the AV. In addition, the AV in the MNS scheme focuses on the number of road sections, where both the driving time and the compensation price are not considered to obtain high utility.
\begin{figure}[t]
\centering
\includegraphics[width=8cm]{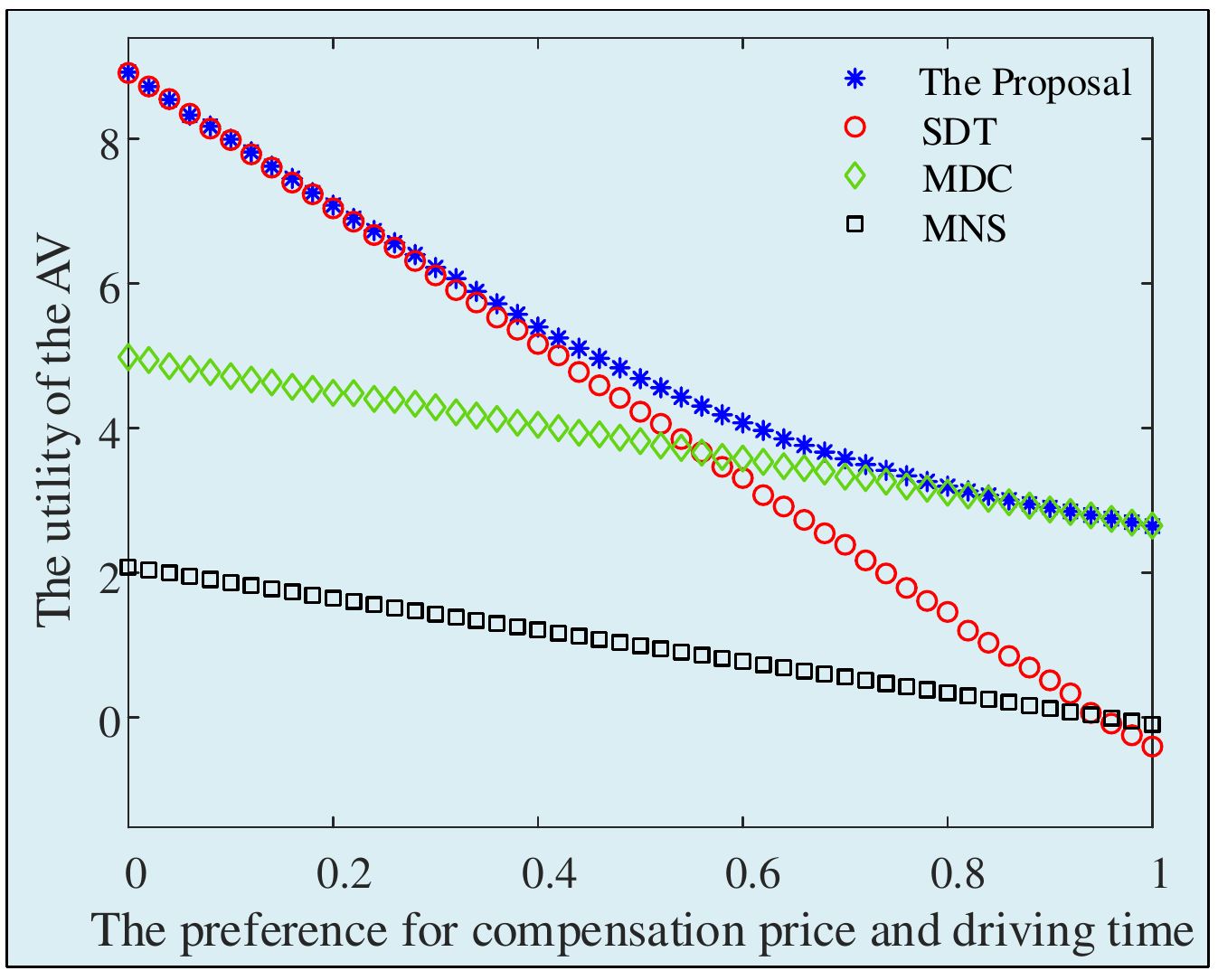}
\vspace{0.0cm}
\caption{The simulation results.}
\label{fig.6}
\end{figure}
\section{Emerging Research Issues}\label{sec:6}
\subsection{High Performance Sensing}\label{sec:41}
To perceive the complex and dynamic traffic information around an AV, it needs to be equipped with sensing devices with high precision. At present, the performance of the sensing devices, however, have constraints in different aspects. For example, the lidar has high resolution and strong anti-interference ability, while its price is very expensive. In addition, the GPS must be used in a non-closed environment with the result that it can not be used in many scenarios, such as tunnels. Therefore, more advanced sensors and the efficient collaboration among sensors need to be further developed to support the FITS.

\subsection{Collaborative Networking, Computing and Caching}\label{sec:43}
With the rapidly development of the sixth generation (6G) networks, the 6G HetVNETs is no longer a simple communication network, but an information network integrating communication, computing, and storage. On the one hand, infrastructures or devices in different layers are different in transmission, computing and caching capabilities. On the other hand, users in the FITS also have diverse requirements for different services. Therefore, the heterogeneous networking, computing and caching resources in the FITS should be well organized and allocated  with the target of satisfying various requirements of services and providing AVs with the customized services.

\subsection{AI-based Algorithms}\label{sec:42}
AI-based algorithms can significantly improve the AV's ability to detect the surrounding objects and facilitate the fusion of data obtained by different sensors. In addition, by analyzing the massive data generated by transportation through AI-based algorithms, the in-depth understanding of road traffic demand and network state can be provided to decision-makers from different network layers. However, different AI-based algorithms are suitable for different applications and require different training time. Therefore, how to design the optimal AI-based algorithm for all the vehicular applications supported by the FITS should be further studied.

\subsection{Network Security}\label{sec:45}
The FITS will generate massive amounts of data to guide the efficient operation of the entire system. In other words, the authenticity and reliability of data are the key factors affecting the efficiency of the FITS. Therefore, how to detect the authenticity and reliability of data in a dynamically evolving FITS is a topic worthy of research. Furthermore, with the designed electronic trading system, a large volume of transactions will occur among network nodes. 1) The system needs to set different rewards for different road sections to dispatch AVs. 2) Rewards should be paid to the nodes in the system which share storage and computing resources. 3) The AVs occupying public facilities need to pay for their requested services. However, the FITS is an open system, some malicious or aggressive nodes may launch attacks to seek personal benefits, which will bring unpredictable losses to normal AVs. Therefore, a green and safe environment needs to be established to solve the security problems in FITS.

\section{Conclusion}\label{sec:7}

The traffic in the city is getting worse with the rapidly increase of vehicles. In order to break the operation mode of the traditional ITS, in this article, we establish a framework of FITS to provide people with safe, efficient and comfortable travel experience. Specifically, we have proposed the network architecture of AVNs based on the rapidly development of AVs and HetNETs. After that, by using the IDT to integrate AVNs and AI, we have designed the FITS to set up an accurate and efficient global traffic scheduling and management system. After that, a case is studied to evaluate the performance of the proposed FITS. Finally, the emerging issues are discussed to highlight the future research directions.

\bibliographystyle{IEEETran}
\bibliography{ref}

\begin{thebibliography}{10}
\providecommand{\url}[1]{#1}
\csname url@samestyle\endcsname
\providecommand{\newblock}{\relax}
\providecommand{\bibinfo}[2]{#2}
\providecommand{\BIBentrySTDinterwordspacing}{\spaceskip=0pt\relax}
\providecommand{\BIBentryALTinterwordstretchfactor}{4}
\providecommand{\BIBentryALTinterwordspacing}{\spaceskip=\fontdimen2\font plus
\BIBentryALTinterwordstretchfactor\fontdimen3\font minus
  \fontdimen4\font\relax}
\providecommand{\BIBforeignlanguage}[2]{{%
\expandafter\ifx\csname l@#1\endcsname\relax
\typeout{** WARNING: IEEEtran.bst: No hyphenation pattern has been}%
\typeout{** loaded for the language `#1'. Using the pattern for}%
\typeout{** the default language instead.}%
\else
\language=\csname l@#1\endcsname
\fi
#2}}
\providecommand{\BIBdecl}{\relax}
\BIBdecl

\bibitem{Rizvi2018aspire}
S.~R. Rizvi, S.~Zehra, and S.~Olariu, ``Aspire: An agent-oriented smart parking
  recommendation system for smart cities,'' \emph{IEEE Intelligent
  Transportation Systems Magazine}, vol.~11, no.~4, pp. 48--61, Winter 2019.

\bibitem{Silva2019information}
F.~A. {Silva}, A.~{Boukerche}, T.~R. M.~B. {Silva}, E.~{Cerqueira}, L.~B.
  {Ruiz}, and A.~A.~F. {Loureiro}, ``Information-driven software-defined
  vehicular networks: Adapting flexible architecture to various scenarios,''
  \emph{IEEE Vehicular Technology Magazine}, vol.~14, no.~1, pp. 98--107, Mar.
  2019.

\bibitem{national}
National Highway Traffic Safety Administration (NHTSA), Critical Reasons for
  Crashes Investigated in the National Motor VehicleCrash Causation Survey,
  2015.

\bibitem{Bhat2018tools}
A.~{Bhat}, S.~{Aoki}, and R.~{Rajkumar}, ``Tools and methodologies for
  autonomous driving systems,'' \emph{Proceedings of the IEEE}, vol. 106,
  no.~9, pp. 1700--1716, Sept. 2018.

\bibitem{Liu2018intelligent}
J.~{Liu} and J.~{Liu}, ``Intelligent and connected vehicles: Current situation,
  future directions, and challenges,'' \emph{IEEE Communications Standards
  Magazine}, vol.~2, no.~3, pp. 59--65, Sept. 2018.

\bibitem{Cheng2020acompre}
N.~{Cheng}, W.~{Quan}, W.~{Shi}, H.~{Wu}, Q.~{Ye}, H.~{Zhou}, W.~{Zhuang},
  X.~{Shen}, and B.~{Bai}, ``A comprehensive simulation platform for
  space-air-ground integrated network,'' \emph{IEEE Wireless Communications},
  vol.~27, no.~1, pp. 178--185, Feb. 2020.

\bibitem{MacHardy2018v2x}
Z.~{MacHardy}, A.~{Khan}, K.~{Obana}, and S.~{Iwashina}, ``V2x access
  technologies: Regulation, research, and remaining challenges,'' \emph{IEEE
  Communications Surveys \& Tutorials}, vol.~20, no.~3, pp. 1858--1877, Feb.
  2018.

\bibitem{Hui2020reservation}
Y.~{Hui}, Z.~{Su}, T.~H. {Luan}, and C.~{Li}, ``Reservation service: Trusted
  relay selection for edge computing services in vehicular networks,''
  \emph{IEEE Journal on Selected Areas in Communications}, vol.~38, no.~12, pp.
  2734--2746, Dec. 2020.

\bibitem{Peng2018vehicular}
H.~Peng, L.~Liang, X.~Shen, and G.~Y. Li, ``Vehicular communications: A network
  layer perspective,'' \emph{IEEE Transactions on Vehicular Technology},
  vol.~68, no.~2, pp. 1064--1078, Feb. 2019.

\bibitem{Shi2018drone}
W.~Shi, H.~Zhou, J.~Li, W.~Xu, N.~Zhang, and X.~Shen, ``Drone assisted
  vehicular networks: Architecture, challenges and opportunities,'' \emph{IEEE
  Network}, vol.~32, no.~3, pp. 130--137, May 2018.

\bibitem{Su2018distributed}
Z.~Su, Y.~Hui, and T.~H. Luan, ``Distributed task allocation to enable
  collaborative autonomous driving with network softwarization,'' \emph{IEEE
  Journal on Selected Areas in Communications}, vol.~36, no.~10, pp.
  2175--2189, Oct. 2018.

\bibitem{Zhao2020Intelligent}
L.~{Zhao}, G.~{Han}, Z.~{Li}, and L.~{Shu}, ``Intelligent digital twin-based
  software-defined vehicular networks,'' \emph{IEEE Network}, vol.~34, no.~5,
  pp. 178--184, Oct. 2020.

\bibitem{Gyawali2020machine}
S.~{Gyawali}, Y.~{Qian}, and R.~Q. {Hu}, ``Machine learning and reputation
  based misbehavior detection in vehicular communication networks,'' \emph{IEEE
  Transactions on Vehicular Technology}, vol.~69, no.~8, pp. 8871--8885, Aug.
  2020.

\bibitem{Zhang2018vehicular}
S.~Zhang, J.~Chen, F.~Lyu, N.~Cheng, W.~Shi, and X.~Shen, ``Vehicular
  communication networks in the automated driving era,'' \emph{IEEE
  Communications Magazine}, vol.~56, no.~9, pp. 26--32, Sept. 2018.

\bibitem{Hui2019agame}
Y.~{Hui}, Z.~{Su}, T.~H. {Luan}, and J.~{Cai}, ``A game theoretic scheme for
  optimal access control in heterogeneous vehicular networks,'' \emph{IEEE
  Transactions on Intelligent Transportation Systems}, vol.~20, no.~12, pp.
  4590--4603, Dec. 2019.

\end{thebibliography}

%\subsection{Results Analysis}

\section{Biography}

Yilong Hui received the Ph.D. degree in control theory and control engineering from Shanghai University, Shanghai, China, in 2018. He is currently a lecturer with the State Key Laboratory of Integrated Services Networks, and with the School of Telecommunication Engineering, Xidian University, China. He has published over 30 scientific
articles in leading journals and international conferences. His research interests include wireless communication, vehicular networks, intelligent transportation systems and autonomous driving. He was the recipient of the Best Paper Award of International Conference WiCon2016 and IEEE Cyber-SciTech2017.
\\

Zhou Su (Corresponding author) is an Associate Editor of IEEE Internet of Things Journal,  IEEE Open Journal of Computer Society, and IET Communications. He is the Chair of the Multimedia Services and Applications over Emerging Networks Interest Group (MENIG) of the IEEE Comsoc Society, the Multimedia Communications Technical Committee. His research interests include multimedia communication, wireless communication and network traffic. He received the best paper award of  IEEE ICC2020, IEEE BigdataSE2019, IEEE ComSoc GCCTC2018, IEEE CyberSciTech2017, etc.
\\

Tom H. Luan is a professor with the School of Engineering, Xidian University, Xi'an, China. He received the B.E. degree from the Xi'an Jiaotong University, China, in 2004, the Master degree from the Hong Kong University of Science and Technology, Hong Kong, in 2007, and the Ph.D. degree from the University of Waterloo, Canada, in 2012, all in Electrical and Computer Engineering. Dr. Luan's research mainly focuses on the content distribution and media streaming in the vehicular ad hoc networks and peer-to-peer networking, protocol design and performance evaluation of wireless cloud computing and fog computing. Dr. Luan has authored/coauthored around 60 journal papers and 30 technical papers in conference proceedings, and awarded one US patent. He served as a TPC member for IEEE Globecom, ICC, PIMRC and the technical reviewer for multiple IEEE Transactions including TMC, TPDS, TVT, TWC and ITS.
\\

Nan Cheng received the Ph.D. degree from the Department of Electrical and Computer Engineering, University of Waterloo in 2016, and B.E. degree and the M.S. degree from the Department of Electronics and Information Engineering, Tongji University, Shanghai, China, in 2009 and 2012, respectively. He is currently a professor with State Key Lab of ISN, and with School of Telecommunication Engineering, Xidian University, Shaanxi, China. He worked as a Post-doctoral fellow with the Department of Electrical and Computer Engineering, University of Toronto, from 2017 to 2018. His current research focuses on space-air-ground integrated system, big data in vehicular networks, and selfdriving system. His research interests also include performance analysis, MAC, opportunistic communication, and application of AI for vehicular networks.

\end{document}